\documentstyle[preprint,aps]{revtex}
\begin{document}
        \tighten
        \title{Measurements of the Total Cross Section for the Scattering
                of Polarized Neutrons from Polarized $^3$He}
        \author{C.~D.~Keith,\footnote{Present address: Indiana
            University Cyclotron Facility, Bloomington, IN 47408}
                        C.~R.~Gould, D.~G.~Haase and M.~L.~Seely} 
        \address{North Carolina State University, Raleigh, NC
                        27695, USA and Triangle Universities Nuclear
                        Laboratory, Durham, NC 27708, USA}
        \author{P.~R.~Huffman,\footnote{Present address: Physics
                        Department, Harvard University, Cambridge, MA
                        02138} N.~R.~Roberson, W.~Tornow and W.~S.~Wilburn}
        \address{Duke University, Durham, NC 27708, USA and Triangle 
                        Universities Nuclear Laboratory, Durham, NC 27708, USA}
%        \date{\today}
        \maketitle
        \widetext

        \begin{abstract}
        Measurements of polarized neutron---polarized $^{3}$He
        scattering  are reported. The target consisted of 
        cryogenically-polarized solid $^{3}$He, thickness 0.04 atom/b 
        and polarization $\sim 0.4$.  Polarized neutrons were produced
        via the
        $^3$H($\vec p$,$\vec n$)$^3$He or $^2$H($\vec d$,$\vec n$)$^3$He
        polarization-transfer reactions.       
        The longitudinal and transverse
        total cross-section differences $\Delta\sigma_L$
        and $\Delta\sigma_T$ were measured for incident neutron energies 
        2--8~MeV\@.  The results are  compared to phase-shift predictions
        based on four different  analyses of  $n$-$^{3}$He scattering.  
        The best agreement is obtained  with a recent $R$-matrix analysis
        of $A=4$ scattering and reaction data, lending strong support
        to the $^4$He level scheme obtained in that analysis.
        Discrepancies with other phase-shift
        parameterizations of $n$-$^3$He scattering exist, attributable 
        in most instances to one or two particular partial waves.
        \end{abstract}

%        \pagebreak
        \narrowtext

\section{Introduction}
        Recent computational advances in the field of 
        few-nucleon dynamics have fueled renewed interest in the 
        three-nucleon and four-nucleon systems \cite{Glo95}.  
        Exact bound-state calculations 
        utilizing realistic, meson-exchange forces are now possible 
        for both the 3N and 4N systems.  Similar calculations
        are currently available for the 3N continuum, and extension 
        to the 4N continuum is under active investigation.
           
        The continuum calculations may prove especially
        revealing, because here exists a fundamental difference 
        between the three- and four-nucleon
        systems:  three-nucleon systems have no excited states
        whereas the four-nucleon system has many. 
        These states (resonances) may
        exhibit sensitivity to the dynamics of the
        nucleon-nucleon interaction and their modification in the
        presence of the nuclear medium.  However, confirming the
        existence, and determining the quantum
        numbers of these resonances is a challenging experimental problem.
        The level scheme proposed as part of a recent
        review article \cite{Til92} has 15 levels at
        excitations 20--30 MeV above the ground state.  For the most part
        these resonances do not appear as sharp structure in any
        scattering or reaction observable,
        and polarization measurements are essential for determining 
        the scattering amplitudes.
       
        While there have been many studies of single polarization observables 
        in 4N systems, there is very little data with
        {\em both\/} polarized 
        target {\em and\/} polarized beam.  Only two measurements have 
        been reported. In 1966,
        Passell and Schermer measured the transmission
        of polarized thermal neutrons through a polarized $^3$He target 
        \cite{Pas66}, and more recently, Alley and Knutsen studied 
         $p$-$^3$He spin-correlation data \cite{All93}.
        In the first experiment, the thermal cross section is completely
        dominated by a single ($0^+$) subthreshold resonance
        in the $^4$He compound nucleus, and no information on the 
        higher-energy resonances was obtained.  The second experiment
        covered a much broader region in the $A=4$ continuum,
        but was sensitive
        only to the isotriplet scattering states.  No previous
        polarized target---polarized beam experiment has fully
        explored a wide range of 4N excited states.
        
        In this paper we report measurements of the longitudinal and 
        transverse neutron total cross-section
        differences $\Delta\sigma_L$ and $\Delta\sigma_T$. These two spin
        observables are directly related to the forward
        elastic-scattering amplitude through the optical theorem
        \cite{Kei94}.
        As such, they allow for a simple interpretation in terms of 
        the properties of the scattering states. The measurements
        were performed at energies corresponding to excitations 22--27~MeV 
        above the  $^4$He ground state, where a number of broad, 
        negative-parity levels are believed to exist. 

        The remainder of this paper is organized as follows.  In 
        Sec.\ \ref{sec:theory} we discuss the basic principles behind
        the measurements.  The 
        polarized target and polarized beam are described in
        Sections \ref{sec:target} and \ref{sec:beam}, respectively.
        The experimental procedure and method of data analysis
        are described in Sec.\ \ref{sec:procedure}.
        The $\Delta\sigma_L$ and $\Delta\sigma_T$ results are
        presented in Sec.\ \ref{sec:results} and compared
        to four separate sets of $n$-$^3$He phase shifts in Sec.\
        \ref{sec:comparison}.  Our conclusions are summarized in Sec.\
        \ref{sec:conclusion}.  A preliminary version of the
        $\Delta\sigma_T$ results has appeared elsewhere~\cite{Kei95b}.
\section{Theory}
\label{sec:theory}
        The formalism necessary to describe the neutron total cross
        section for polarized target and polarized beam has been developed and
        discussed in an earlier paper~\cite{Kei94}.
        For the sake of clarity we briefly review the results of that
        paper in the following section.

        The parity-conserving,
        time-reversal invariant part of the forward elastic scattering
        amplitude allows the neutron total cross section for spin-1/2
        nuclei to be expressed as
\begin{equation}
        \sigma_{tot} = \sigma_{0} + \frac{1}{2}\Delta\sigma_L
        P^{z}_{n}P^{z}_{t} +
        \frac{1}{2}\Delta\sigma_T(P^{x}_{n}P^{x}_{t} + P^{y}_{n}P^{y}_{t}).
                                                        \label{sigtota}
\end{equation}
        Here $\sigma_{0}$ is the unpolarized neutron total cross
        section, $P^i_n$ ($P^i_t$) is the $i$th projection of the beam
        (target) polarization axis,
        and $\Delta\sigma_L$ and $\Delta\sigma_T$ are the        
        longitudinal and transverse total cross-section differences.
        The latter may be represented pictorially as\footnote{Note that
        our definition of parallel minus antiparallel
        is exactly opposite to the convention commonly used for 
        nucleon-nucleon scattering.}
\begin{equation}
                \Delta\sigma_L=\sigma_{tot}(\stackrel{\textstyle\rightarrow}
                        \rightarrow) 
                        -\sigma_{tot}(\stackrel {\textstyle\rightarrow}
                         \leftarrow)                          \label{arrowsL}
\end{equation}
        and 
\begin{equation}
                \Delta\sigma_T=\sigma_{tot}(\uparrow\uparrow) 
                        -\sigma_{tot}(\uparrow \downarrow).    \label{arrowsT}
\end{equation}
        The measurements of $\Delta\sigma_L$ and
        $\Delta\sigma_T$ reported here were performed separately,
        with both beam and target polarized along axes longitudinal to
        the incident beam direction (i.e. along the $z$ axis) or transverse
        to the beam direction (i.e. along the $y$ axis).  Hence in 
        subsequent discussion, the superscripts used to indicate
        the polarization axes will be dropped.

        An accurate determination of the $\Delta\sigma_L$ and
        $\Delta\sigma_T$ observables does
        not require an absolute measurement of the total cross section.
        Rather, the cross-section differences may be extracted from
        transmission measurements of polarized neutrons through
        the polarized target.
        The attenuation of an incident beam of $N_0$ polarized
        neutrons due to a polarized target of areal density $\tau$
        will be
\begin{equation}
        N_{\pm}/N_0 = \exp \left[ - \tau(\sigma_0 \pm 
                                   \frac{1}{2}\Delta\sigma_L P_nP_t) \right]
                                                        \label{attenuateL}
\end{equation}
        when both the beam and target are longitudinally polarized, or
\begin{equation}
        N_{\pm}/N_0 = \exp \left[ - \tau(\sigma_0 \pm
                                   \frac{1}{2}\Delta\sigma_T P_nP_t) \right]
                                                        \label{attenuateT}
\end{equation}
        when beam and target are polarized in the transverse
        direction.  The $\pm$ signs are used to
        indicate whether the beam and target spins are polarized
        parallel ($+$) or antiparallel ($-$) to one another.
        By periodically reversing the spin of
        the beam (or target), one may observe an asymmetry in the
        attenuation of the beam.
        This neutron-transmission or spin-spin asymmetry is defined as
\begin{eqnarray}
        \varepsilon_{L,T} &=& \frac{N_{+} - N_{-}}{N_{+} + N_{-}}
                                        \label{ntasym}  \\
                  &=& \tanh \left[ -\frac{1}{2} \Delta\sigma_{L,T}
                                  P_n P_t \tau \right]  \nonumber \\
                          &\approx& -\frac{1}{2} \Delta\sigma_{L,T}
                                            P_n P_t \tau. \nonumber
\end{eqnarray}
        Because the observed values of 
        $\varepsilon$ are typically of order $10^{-3}$, the error 
        introduced by replacing the hyperbolic tangent with its 
        argument in Eq.~(\ref{ntasym}) is negligible.  In
        writing Eq.~(\ref{ntasym}) we have assumed
        that the incident neutron flux $N_0$ is unaffected by reversing
        the spin.
        
        Because they are directly related to the forward
        scattering amplitude via the optical theorem,
        $\sigma_0$, $\Delta\sigma_{L}$, and $\Delta\sigma_T$
        are not sensitive
        to interference effects between the various partial waves.
        In principle this allows for a much simpler interpretation
        of the scattering and reaction processes.
        The $n$-$^3$He total
        cross sections can be expressed as linear sums of
        ``partial-wave'' cross sections $\sigma(J,l,s,l',s')$, where
\begin{equation}
        \sigma(J,l,s,l',s') = {\rm Re}\left\{\frac{\pi}{2k^2}
                        (2J+1)
                          \left[\delta_{ll'}\delta_{ss'} - 
                            S^{J}_{ll'ss'}\right]\right\}.
                                                       \label{partxsect}
\end{equation}
        Here $J,l,s$ ($J,l',s'$) are the incoming (outgoing)
        total, orbital, and channel-spin angular momenta, $k$ 
        is the neutron center-of-mass
        wave number, and $S^{J}_{ll'ss'}$ is the elastic 
        scattering matrix element
        that describes transitions from the initial neutron channel
        ($J,l,s$) to the final neutron channel ($J,l',s'$).
        For partial waves up to $l=1$, we find
\begin{eqnarray}
        \sigma_0 &=& \sigma(^1\!S_0) + \sigma(^3\!S_1) +
                         \sigma(^1\!P_1)         \nonumber\\
                 & & + \sigma(^3\!P_0) + \sigma(^3\!P_1) + \sigma(^3\!P_2),
                                                     \label{sig0}
\end{eqnarray}
\begin{eqnarray}
        \Delta\sigma_L &=& -2\sigma(^1\!S_0) + \frac{2}{3}\sigma(^3\!S_1)
                       - 2\sigma(^1\!P_1) \nonumber \\
                       & & -2\sigma(^3\!P_0) + 2\sigma(^3\!P_1) + 
                       \frac{2}{5}\sigma(^3\!P_2) \nonumber \\
                       & & + \frac{4}{3}\sqrt{2}\sigma(^3\!S_1-^3\!\!D_1),
                                                     \label{DsigL}
\end{eqnarray}
        and
\begin{eqnarray}
        \Delta\sigma_T &=& -2\sigma(^1\!S_0) + \frac{2}{3}\sigma(^3\!S_1)
                       - 2\sigma(^1\!P_1) \nonumber \\
                       & & +2\sigma(^3\!P_0) + \frac{4}{5}\sigma(^3\!P_2)
                       - \frac{2}{3}\sqrt{2}\sigma(^3\!S_1-^3\!\!D_1).
                                                     \label{DsigT}
\end{eqnarray}
        A negative coefficient
        simply implies that the cross section for this particular
        wave is greater when the beam and target spins are
        antiparallel to one another.

        Detailed predictions of
        the total cross-section differences $\Delta\sigma_L$ and
        $\Delta\sigma_T$ for $\vec{n}$-$\vec{^3{\rm He}}$ scattering
        were presented in an earlier paper \cite{Kei94}.
        These calculations were based
        on three separate analyses of $n$-$^3$He  scattering
        and reaction data \cite{Hal89,Jan88,Lis75}, as well as 
        a microscopic resonating-group model calculation of
        the 4N excited states \cite{Hof93}.  It was observed that,
        although the different sets of phase shifts
        provide adequate descriptions of pre-existing data, 
        they predict quantitatively different
        values of both $\Delta\sigma_L$ and $\Delta\sigma_T$ in the present
        region of interest.  In most instances these discrepancies could
        be attributed to one or two partial waves.  Based on this
        observation, we concluded that comprehensive measurements of both
        $\Delta\sigma_L$ and  $\Delta\sigma_T$, in 
        combination with the unpolarized neutron total cross section
        $\sigma_0$, could be used to extract specific
        information about the
        partial-wave content of the resonating $^4$He compound
        nucleus.
\section{Experimental Apparatus}
\subsection{Polarized $^3$H{e} Target}
   \label{sec:target}
        A brief description of the polarized solid $^3$He target is
        given below.  The target is described in greater detail in
        Ref.~\cite{Kei95}.  

        Owing to the low intensity of the
        polarized neutron beam, an extremely thick sample of polarized
        $^3$He is desirable.
        In a measurement of the neutron transmission asymmetry 
        $\varepsilon$, the number of observed neutron counts $N$ necessary
        to obtain a statistical precision
        $\Delta\varepsilon/\varepsilon$ is
\begin{equation}
                N = \frac{1}{2}[\frac{\Delta \varepsilon}{\varepsilon}
                      \Delta\sigma P_n P_t \tau]^{-2}.  \label{precision} \\
\end{equation}
        The factor of one-half means that $N$ counts are needed in both the
        {\it up} and {\it down\/} spin states. 
        This indicates that, for given values of $P_n$ and $\Delta\sigma$,
        the figure of merit for comparing
        polarized targets in transmission experiments
        should be $\tau^2 \: P_t^2$.  The figure of merit for the
        TUNL solid $^3$He target exceeds current polarized $^3$He
        gas targets by nearly two orders of magnitude.  Furthermore,
        the densities of the 
        condensed phases of $^3$He correspond to nearly 100~MPa 
        of room temperature gas, while targets of polarized $^3$He gas
        are limited to 1~MPa.

        The liquid phase of $^3$He behaves as a Fermi liquid and
        can not be polarized to any great extent.  On the other
        hand, solid $^3$He is a nuclear
        paramagnet and can be polarized by the static or ``brute-force'' 
        method; the sample is cooled to a very low ($\sim 10$~mK)
        temperature in the presence of an externally-applied magnetic field 
        ($\sim 7$~T).
        The resulting polarization for the body-centered cubic (bcc)
        phase at temperature $T$ and field $B$ is given by the
        Brillouin expression
\begin{equation}
                P_t = \tanh\left[\frac{1}{k_B T}(\mu B + \Theta P_t + 
                                KP_t^3)\right].         \label{Brill}
\end{equation}
        where $\mu=-2.13 \mu_{\rm N}$ is the magnetic moment of the
        $^3$He nucleus, and $k_B$ is Boltzmann's constant.  
        The quantities $\Theta$ and $K$
        are corrections to the Curie law of paramagnetism and
        describe the anti-ferromagnetic 
        exchange of neighboring $^3$He atoms in the bcc lattice.
        Therefore,
        the actual polarization of solid $^3$He is slightly lower
        than that calculated assuming simple paramagnetic behavior.
        Values for these corrections ($\Theta/k_B = -1.18$~mK and 
        $K/k_B=-1.96$~mK) were determined
        by fitting the observed low-temperature properties of bcc solid
        $^3$He \cite{Sti85a}.

        For the measurements reported in this paper, a $^3$He-$^4$He dilution 
        refrigerator was used to cool the sample to approximately 12~mK in 
        an externally-applied magnetic field of 7~T\@.  The field was 
        provided
        by a superconducting split-coil magnet operated in
        persistent-current mode.  The magnet was physically rotated to 
        provide fields either parallel (longitudinal) or perpendicular
        (transverse) to the incident beam direction.
        The lowest target temperature obtained during these
        measurements was $11.9 \pm 0.2$~mK,
        corresponding to $38.7 \pm 0.6$\% polarization.

        The sample cell for the target
        is shown in Fig.~\ref{fig:target}.  The cylindrical container was
        constructed primarily of beryllium copper (BeCu) with four
        flat surfaces
        machined from the cylinder.  The flats reduced the amount of material
        to be cooled and minimized the attenuation of the neutron
        beam due to BeCu.  The wall thickness at the flats was 1.27~mm
        perpendicular to the beam and 2.54~mm parallel to the beam.
        The sample cell was thermally anchored to the dilution refrigerator's
        mixing chamber by a 45~cm long OFHC copper cold finger.
        
        The cell was filled with $^3$He through a 0.75~mm
        I.D. stainless steel tube hard-soldered into the top of the
        cell.  Cupro-nickel capillary (0.1~mm I.D.) connected the fill
        tube to a room temperature gas-handling system.
        The interior sample space was a rectangular parallelepiped with
        dimensions $38.1 \times 14.0 \times 21.6$~mm and was filled with
        3~micron silver
        powder packed to 19\% of the density of solid silver.  The powder was
        used to provide good thermal contact between the solid $^3$He and the 
        BeCu cell, ensuring a homogeneous temperature throughout the target.
        
        The solid $^3$He sample was grown by first filling the cell
        with liquid
        $^3$He at approximately 3~K\@.  The liquid was then compressed to a 
        density of  0.125~g/cm$^3$ by increasing the $^3$He vapor 
        pressure to 3.6~MPa.
        At this density solid began to form at 1.1~K, and the sample was
        completely solidified at 0.83~K \cite{Gri71}.
        With the silver powder in place, the thickness of the solid $^3$He
        sample was $4.34\pm 0.09 \times 10^{22}$~atoms/cm$^2$.

        The target polarization was extracted from the
        temperature of the BeCu sample cell, as measured by two
        independent thermometric standards: a $^{60}$Co\underline{Co}
        nuclear orientation thermometer \cite{Mar83} and
        a $^3$He melting curve thermometer (MCT) \cite{Kei92}.
        The nuclear orientation thermometry required
        an intrinsic germanium detector to observe
        the 1.17~MeV and 1.33~MeV $\gamma$~rays from $^{60}$Co.
        To avoid radiation damage from neutrons however, the detector had to
        be removed
        from the experimental hall whenever beam was on target.  Therefore
        the $^{60}$Co\underline{Co} measurements were made immediately
        before and
        after each neutron asymmetry measurement.  The melting curve
        thermometer on the other hand, could be used throughout
        the neutron measurements.
        The output of the MCT was read directly into the data acquisition 
        computer and the temperature sampled every 100~ms.
        The average polarization of the target
        for a particular asymmetry measurement
        was determined from the average MCT temperature during that time.
        With no beam on target the $^{60}$Co\underline{Co} 
        and MCT were found to agree
        within $\pm$2\%, and although the MCT could resolve
        temperature changes as small as one
        microkelvin, no significant warming due to neutron or 
        $\gamma$-ray interactions within the target was observed.
        
\subsection{The Polarized Beam}
      \label{sec:beam}
    \subsubsection{Neutron Production and Detection}
        Polarized neutrons were produced as secondary beams from either the
        $^3$H($\vec p$,$\vec n$)$^3$He or $^2$H($\vec d$,$\vec n$)$^3$He
        polarization-transfer reactions at
        $0^{\circ}$.  The $^3$H($\vec p$,$\vec n$)$^3$He 
        reaction was used to produce neutrons
        with energy less than 4~MeV because it has a negative Q-value,
        -0.764~MeV\@.  However, safety considerations limited the maximum
        amount of tritium that could be used, and the resulting neutron fluxes
        were low. The $^2$H($\vec d$,$\vec n$)$^3$He reaction (Q=3.269~MeV)
        was used at higher energies, and the neutron fluxes here were
        typically 10--20 times greater.

        The polarized charged-particle beams were produced by the TUNL
        atomic beam polarized ion source \cite{Cle95a,Cle95b,Din95} 
        and accelerated by a 
        tandem Van de Graaff.  The ion
        source produced a polarized beam whose spin axis was parallel
        to its momentum.  A Wien filter
        located between the source and accelerator was used to rotate the
        polarized beam's spin axis to the desired orientation, longitudinal
        or transverse, at the neutron production target. 
        The position of the beam was feedback-stabilized in both the 
        horizontal and vertical planes by four sets of steering magnets.
        Computer-controlled steering was used to maintain the beam
        position at the center of a rotating-wire scanner
        installed inside the beam pipe
        approximately 2~m from the neutron-production target.

        The $^3$H($\vec p$,$\vec n$)$^3$He neutron production target
        was a  tritiated-titanium foil, backed by a 0.51~mm thick copper
        disk. A 0.1~MPa $^4$He
        gas cell, with a 2.54~$\mu$m Havar entrance window,
        surrounded the tritiated foil to prevent contamination of 
        the beam line.
        The $^2$H($\vec d$,$\vec n$)$^3$He neutron-production 
        target was a deuterium gas cell, 60~mm long, 19~mm in
        diameter and  operated at a D$_2$ pressure of 0.4~MPa.  The Havar
        window for this cell was 6.35~$\mu$m thick, and the
        deuteron beam was stopped by a 0.51~mm tantalum disk.  Both
        neutron-production targets were air cooled.  To eliminate the
        deflection of the charged particles due to the superconducting 
        magnet, the last 1.2~m of beam pipe was constructed of soft
        iron and lined with a high permeability iron-nickel alloy.
     
        The neutron-production targets were located as close as
        possible to the polarized target.  Neutron collimation and
        detector shielding were accomplished
        by a combination  of copper and polyethylene as shown
        in Fig.~\ref{fig:poltar}.  The copper preshield located
        between the neutron production target and polarized target
        reduced the number of neutrons striking the superconducting
        magnet.  The polyethylene collimation system located after
        the polarized target defined a 25.7 $\times$ 9.4~mm beam spot
        at the center of the polarized target, corresponding to a
        solid angle of approximately 0.5~msr.
        
        Neutrons that were transmitted through the polarized target
        were detected by two liquid scintillators located at 
        $0^{\circ}$ and surrounded by a polyethylene shield.
        The scintillation liquid (BC501) was contained in two
        cylindrical aluminum containers (127~mm diameter, 127~mm long),
        each with an optically transparent endcap 
        coupled to a 127~mm diameter photomultiplier tube.
        The cylinders were placed one atop the other, with their
        axes, as well the photomultiplier tubes, pointing in the
        vertical direction.
        Pulse-shape discrimination (PSD) was performed on the phototube
        anode signals to distinguish neutron events from $\gamma$ rays.
        The PSD was performed by commercially-manufactured modules
        \cite{PSD95}, with pulse-height thresholds set to discriminate
        against low-energy neutrons.
        Valid neutron events were counted in scalers, 
        and stored in the computer at set intervals.  
        
        The collimation/detection system was tested in two ways. 
        First, the alignment of the target and collimator was verified
        by exposing x-ray films to the gammas produced by the 
        charged-particle beam.  This showed that the target completely
        filled the acceptance angle of the collimator.
        Second, neutron time-of-flight measurements were performed 
        with a pulsed beam at 10 MeV. The time-of-flight spectrum
        showed that only neutrons of the correct energy were being
        counted.  In addition, blocking the exit of the collimator 
        with 30~cm of tungsten followed by 30~cm of polyethylene
        reduced the neutron count rate by a factor of $10^3$,
        indicating that the detectors were adequately shielded from 
        energetic background neutrons.

        The neutron transmission
        asymmetries were  observed by  reversing the spin
        of the charged-particle beam every 100 ms.
        For an accurate measurement it is necessary
        to know the ratio of the neutron fluxes produced
        by each spin state of the charged-particle beam.
        For the $^3$H($\vec p$,$\vec n$)$^3$He 
        reaction, the neutron yield is proportional to the 
        proton beam current, and it proved sufficient to count the digitized
        beam current in each spin state, using the ratio 
        to normalize the neutron fluxes.
        Such normalization does not work for the
        $^2$H($\vec d$,$\vec n$)$^3$He reaction because here the yield
        depends on the tensor polarization
        $P_{zz}$ of the deuteron beam as well as on the
        beam intensity.  The polarized source was operated in a manner
        such that, ideally, $P_{zz}$ remained constant while the
        vector polarization was completely reversed.  In practice,
        we determined that the tensor component
        changed by as much as a few percent when the deuteron spin was
        flipped.
        To monitor the flux more reliably, we placed 
        a third liquid scintillator at $0^{\circ}$,
        between the copper preshield and the $\vec{^3{\rm He}}$ target.
        This monitor detector was used to normalize the number of
        neutrons in the {\it up\/} and {\it down\/} spin states 
        to the same incident flux.  Because of its close proximity to
        the superconducting magnet, the monitor detector
        was optically coupled to a 51~mm diameter phototube by a 1~m
        long light pipe.
        Due to its small dimensions ($25.4 \times11.1 \times 22.2$~mm), 
        $\gamma$ rays did not deposit much energy in this detector and could
        be separated from the neutron events by pulse height alone, 
        without the need for pulse-shape discrimination.

        Corrections to the incident-flux normalization described above
        are discussed in Sec.\ \ref{sec:background}.
  \subsubsection{Polarization of the neutron beam}
       The polarization of 
       the charged-particle beam was measured with a
       carbon-foil polarimeter.  The polarization of the  
       neutron beam was then calculated
       from the polarization of the charged particles, using 
       known polarization-transfer coefficients. The polarimeter
       consisted of a thin (5~$\mu$m/cm$^2$) 22~mm
       diameter carbon foil located at the center of a small
       scattering chamber.
       Two silicon surface-barrier detectors detected
       protons from the $^{12}$C($\vec{p},p$)$^{12}$C reaction, or the
       $^{12}$C($\vec{d},p_0$)$^{13}$C reaction.
       The detectors were located at  $\pm 40^{\circ}$, and a 
       tantalum collimation system defined a $\pm3.5^{\circ}$ angular
       acceptance for each.
       The carbon foil was mounted to an 
       aluminum plunger inside the polarimeter and was removed from
       the beam path when not in use.

       To determine the polarization of the proton
       or deuteron beam, a left-right asymmetry was measured between
       the two silicon detectors.  The asymmetry was measured for both the
       {\it up\/} and {\it down\/} spin states and the difference taken 
       to cancel systematic effects.
       The  average beam polarization was calculated on the basis of this
       average polarimeter asymmetry, $\varepsilon_{pol}$:
\begin{equation}
                \varepsilon_{pol} = \frac{1}{2}\left[
                \frac{L^+ - R^+}{L^+ + R^+} - \frac{L^- - R^-}{L^- + R^-} 
                \right].                        \label{polasy}
\end{equation}

        When polarized protons were used to produce neutrons,
        the polarimeter measured a left-right asymmetry for the elastic
        scattering of protons from the carbon foil.
        Published values~\cite{Mos65,Ter68} of the
        $^{12}$C($\vec{p},p$)$^{12}$C analyzing power $A_y$ were then
        used to calculate the average neutron polarization,
\begin{equation}
                P_{n}=\frac{\varepsilon_{pol} C_1 K^{y'}_{y}}{A_{y}}
                                                        \label{neutpol} \\
\end{equation}
        Here $K^{y'}_{y}$ (or $K^{z'}_z$ in the case of a longitudinally 
        polarized beam) is the polarization-transfer coefficient
        for the $^3$H($\vec p$,$\vec n$)$^3$He reaction, and
        $C_1$ is a correction term that describes the depolarization
        of the neutron beam as it passes through the field of the 7~T 
        superconducting magnet.  The effect is small 
        ($C_1 = 0.978$ at 1.94~MeV and $C_1=0.984$ at 3.65~MeV)
        because the dominant field component is parallel to the neutron spin
        axis.  The values of $C_1$ were 
        determined from a detailed calculation of the magnetic field 
        \cite{Wil95}.

        At the lowest proton energy,
        $E_p = 3.0$~MeV, the  $^{12}$C($\vec{p},p$)$^{12}$C
        analyzing power was too small to be useful as a
        polarization monitor.  Therefore all measurements of the
        proton polarization were made at $E_p = 4.7$~MeV\@.

        When deuterons were used to produce the polarized neutron beam,
        $P_n$ was again determined from the polarimeter asymmetry,
\begin{equation}
        P_n = \frac{C_1 \varepsilon_{pol}}{A_{ef\!f}}.   \label{effanal}
\end{equation}
        Here $A_{ef\!f}$ is an ``effective'' analyzing power that
        relates the polarimeter asymmetry measured for the
        $^{12}$C($\vec{d},p_0$)$^{13}$C reaction to the resulting neutron 
        polarization from the $^2$H($\vec d$,$\vec n$)$^3$He reaction.
        This effective analyzing power was measured in a separate 
        experiment with a neutron polarimeter consisting of a $^4$He
        scatterer.   In this experiment the $^{12}$C($\vec{d},p_0$)$^{13}$C
        asymmetries were calibrated against known 
        $\vec{n}$-$^4$He analyzing powers \cite{Tor74}.
        The correction factor due to the 
        superconducting magnetic field, $C_1$, has been described
        above, with $C_1=0.987$ at 4.95~MeV and $C_1=0.990$ at 7.46~MeV\@.

        During the measurements of $\Delta\sigma_T$, 
        the polarimeter asymmetries were measured approximately every 2--3
        hours.  Under normal operation of the polarized ion source,
        we found that the polarimeter asymmetries stayed constant
        (within experimental uncertainties) during the course
        of several days.
        During the $\Delta\sigma_L$ measurements,
        the proton and deuteron polarizations could be measured only
        after the Wien filter was used to rotate their spins perpendicular
        to the beam. This involved retuning the beam optics, and so the 
        measurements were performed only twice at each beam
        energy: immediately before and immediately after the longitudinal
        neutron-transmission asymmetry was measured.
\section{Experimental Procedure and Data Analysis}
\label{sec:procedure}
\subsection{Measurement of Spin-Spin Asymmetries}
\label{sec:spinspin}
        Cooling of the solid $^3$He target commenced approximately 24~hours 
        before measurements of the neutron-transmission asymmetry.
        During this time both the $^{60}$Co\underline{Co} 
        nuclear orientation and $^3$He
        melting curve thermometers were used to monitor the target 
        temperature.
        After the target reached a temperature of 15~mK,
        the germanium detector for the $^{60}$Co\underline{Co}
        thermometer was removed from the experimental hall and the neutron
        measurements began.

        The spin of the neutron beam was reversed every 100~ms
        by toggling radio-frequency transition units
        at the polarized ion source.  The spins were flipped
        according to an eight-step sequence
        (${}+{}-{}-{}+{}-{}+{}+{}-{}$)
        to minimize effects which arise from
        drifts in detector efficiency that are linear or quadratic in time.
        Typical count rates encountered during these measurements were
        between $10^4$~s$^{-1}$ when the $^2$H($\vec d$,$\vec n$)$^3$He
        reaction was used and 400~s$^{-1}$ with the
        $^3$H($\vec p$,$\vec n$)$^3$He reaction.

        The data consisted of CAMAC scaler counts of neutron events, 
        digitized charged-particle beam current,
        digitized polarized target temperature, and events of a
        100~kHz dead-time pulser.
        At the end of each eight-step sequence, a count-down scaler
        was decremented from its preset value of 1024.  Data were
        stored in the computer buffer as spectra of scaler counts versus
        time, each eight-step sequence comprising one channel of the
        various spectra.  When the count-down scaler reached zero,
        acquisition was inhibited, and the data were written to the computer
        disk.  All spectra were then cleared, the count-down scaler
        reset to 1024, and data acquisition recommenced.
        A ``run'' therefore consisted of 1024 eight-step
        sequences and required about fifteen minutes of beam time.
        At each energy the data were collected for
        about twelve hours with a polarized target and for an equal
        amount of time with an unpolarized target.
                
        Two spectra were allocated for each observable of interest,
        one for neutron-spin up (parallel to target spin), the other
        for neutron-spin down (antiparallel to target spin).
        Acquisition into the spin-up spectra was inhibited during
        the spin-down portions of each eight-step sequence and vice
        versa.  All data acquisition
        was inhibited 2~ms prior to and 5~ms after each spin flip to
        give the beam polarization time to stabilize.  Acquisition was also
        halted whenever the beam current fell above or below
        prescribed limits. To ensure that equal time was spent in both
        the up- and down-spin states, the entire eight-step sequence
        during which the beam current had fallen outside its limits
        was rejected in the final
        data analysis.  These occurrences were easily observed in the
        dead-time pulser spectra.  In all, less than 1\% of the
        data were rejected for this reason.

        Transmission asymmetries for the two 
        main neutron detectors were 
        calculated for each eight-step sequence according to
\begin{equation}
                \varepsilon = \frac{\tilde{N}_+ - \tilde{N}_-}
                                   {\tilde{N}_+ + \tilde{N}_-}.  
                                   \label{ntasym2}
\end{equation}
        Here $\tilde{N}_{\pm}$ is the number of dead time-corrected
        neutron counts in each spin state normalized to the incident 
        neutron flux,
\begin{equation}
                \tilde{N}_{\pm} = \frac{N_{\pm}}{I_{\pm}}.
                                                        \label{normcounts}
\end{equation}
        The normalization factor $I$ is either the proton beam current
        or the yield in the neutron monitor detector.
        The transmission asymmetries for all eight-step sequences were 
        combined in a weighted average for both the top and the bottom
        neutron detectors.  These two results were then combined to 
        give the average neutron transmission
        asymmetry $\bar{\varepsilon}$ and its associated statistical
        uncertainty.  A standard deviation was calculated for each 
        set of data, to compare with the standard deviations 
        expected from Poisson counting statistics.
\subsection{Measurement of Background Asymmetries}
\label{sec:background}
        After each measurement of the transmission asymmetry,
        the solid $^3$He target was melted and warmed to 1~K, and
        the neutron transmission
        measurements were repeated. The superconducting magnet
        continued to operate in persistent-current mode.
        The polarization of the liquid phase at 1~K was less than 
        0.5\%, while its density was 8\% less than the density of the solid.
        These measurements were performed to determine how much of the
        observed neutron transmission asymmetry was due to effects 
        other than spin-dependent forces between the
        polarized neutron beam and polarized target nuclei.
        
        While these background, or ``warm'',
        asymmetries were typically
        an order of magnitude lower than the spin-spin
        asymmetries, they were, in general, non-zero and
        were subtracted from the spin-spin, or ``cold'' measurements.  
        As discussed below, the asymmetries observed during the
        warm measurements were due
        to polarization effects associated with the incident
        charged-particle beams.  As long as the beam polarization
        remained constant during the warm and cold asymmetry measurements,
        the background asymmetry was the same in both measurements and
        so the warm asymmetry could simply be subtracted from the cold.
        If the polarization differed between the two
        measurements, a correction based on the two polarizations
        had to be made.  Thus, proper correction for the background 
        asymmetries required some understanding of their origin.

        When the $^3$H($\vec p$,$\vec n$)$^3$He source reaction was used, 
        the incident proton current provided the
        normalization factor.  However, if imperfect 
        alignment exists between the proton beam and neutron collimation, 
        the vector analyzing power of this reaction will
        produce a non-zero asymmetry in incident neutron flux that
        is not eliminated by the beam-current normalization.
        The vector analyzing power for this reaction
        vanishes at $0^{\circ}$ and thus
        the observed background asymmetries (and the subsequent
        corrections) are small.  Since the asymmetry produced by
        the vector analyzing power
        is proportional to the polarization of the proton beam $P_p$,
        the background asymmetry observed
        during the warm measurement is scaled to the same value of
        $P_p$ that existed during the cold measurement.
        Therefore, $\Delta\sigma_T$
        is extracted from the difference between the 
        cold and warm asymmetries, with the latter scaled
        by $P_p$,
\begin{equation}
                \Delta\sigma_T= \frac{-2}{P_t P_n \tau}
                \left[ \bar{\varepsilon}_c -
                       \frac{P_{pc}}{P_{pw}}\bar{\varepsilon}_w \right].
                                                        \label{deltasigT}
\end{equation}
        Here the subscripts $c$ and $w$ refer to the cold
        and warm measurements, and $P_n$ is the value
        of neutron polarization during the cold asymmetry
        measurement.

        Since parity conservation forbids any longitudinal analyzing
        powers for the $^3$H($\vec p$,$\vec n$)$^3$He reaction, 
        no background correction should be necessary for the
        low-energy longitudinal measurements.  The
        warm measurement at 3.65~MeV was in fact consistent 
        with zero.  Thus $\Delta\sigma_L$ (at this energy)
        was determined from the cold asymmetry measurements alone,
\begin{equation}
                \Delta\sigma_L = \frac{-2}{P_t P_n \tau}
                \bar{\varepsilon}_c.
                                                        \label{deltasigL}
\end{equation}

        In the case of the $^2$H($\vec d$,$\vec n$)$^3$He reaction
        we must consider two sources of background asymmetry.
        In addition to a vector analyzing power (which
        produces an asymmetry in the neutron yield at non-zero
        angles), this reaction possesses a tensor
        analyzing power that affects the 
        neutron yield at $0^{\circ}$.  If $I_0$ is the  
        $0^{\circ}$ yield from a
        completely unpolarized deuteron beam, then the yield from a
        polarized beam, $I(0^{\circ})$, will be
\begin{equation}
        I(0^{\circ}) = I_0(1-\frac{1}{4}A_{zz}P_{zz})   \label{yieldT}
\end{equation}
        in the transverse geometry, or
\begin{equation}
        I(0^{\circ}) = I_0(1+\frac{1}{2}A_{zz}P_{zz})   \label{yieldL}
\end{equation}
        in the longitudinal geometry.  Here $P_{zz}$ is the
        longitudinal tensor polarization of the deuteron beam 
        and $A_{zz}$ is the tensor analyzing power for the 
        $^2$H($\vec d$,$n$)$^3$He reaction.
        If there is a change in tensor polarization,
        $\Delta P_{zz}$, when the 
        deuteron spin is flipped at the polarized ion source,
        then an asymmetry in the $0^{\circ}$ neutron yield
        will result.

        To monitor the $0^{\circ}$ yield,
        a thin scintillator was placed between the $^3$He target 
        and neutron production target (see Section~\ref{sec:beam}).
        However, the solid angle subtended
        by the monitor detector was slightly different from that
        of the main detector.  This led to an
        asymmetry in the monitor-normalized neutron counts
        caused by either the vector or tensor analyzing power, or both.
        A vector analyzing power is parity-forbidden
        for a longitudinally-polarized deuteron beam, and during
        the transverse measurements we observed
        little change in the deuteron vector polarization.
        Therefore the only correction necessary for the
        $^2$H($\vec d$,$\vec n$)$^3$He measurements
        was one based on the {\em tensor\/} analyzing power.
        Since both the background asymmetry and the asymmetry observed
        by the monitor detector were proportional to $\Delta P_{zz}$, 
        it proved convenient to use the monitor asymmetry 
        (which was measured with a high degree of
        statistical accuracy) to scale the warm to
        cold asymmetry measurements.  Thus, for the
        $^2$H($\vec d$,$\vec n$)$^3$He measurements,
\begin{equation}
                \Delta\sigma_{L,T}= \frac{-2}{P_t P_n \tau}
                \left[ \bar{\varepsilon}_c -
                       \frac{\varepsilon_{mc}}{\varepsilon_{mw}}
                         \bar{\varepsilon}_w \right],
                                                        \label{deltasigL2}
\end{equation}
        where $\varepsilon_{mc}$ ($\varepsilon_{mw}$) are the cold
        (warm) monitor asymmetries.
 
        According to Eqs. \ref{yieldT} and \ref{yieldL}, the asymmetry
        resulting from a given value of $\Delta P_{zz}$ should be twice
        as large and of the opposite
        sign in the longitudinal geometry as in the transverse.
        This explains why the background
        asymmetries were typically larger (and of the opposite sign)
        during the high-energy measurements of $\Delta\sigma_L$ 
        (Tables~\ref{tab:transasym} and \ref{tab:longasym}).
        
        Additional $\Delta\sigma_T$ measurements were made with a 
        cold ($\sim15$~mK),
        empty sample container, a cold container filled with liquid
        $^3$He, and a warm empty container.
        Such measurements are sensitive
        to spin-spin effects caused by polarizable materials in the sample
        container other than $^3$He (e.g. copper).  With the exception
        of the cold liquid measurements at 1.94 and 3.65~MeV, all such
        background measurements were consistent with the corresponding
        warm, unpolarized measurements.  The asymmetries observed with
        the cold liquid target were in fact consistent with a 
        $^3$He polarization of 3\%, the expected polarization of
        $^3$He in the liquid phase at 12~mK \cite{Kei95,Ram70}.
        At no time did we observe effects due to polarizable 
        materials other than $^3$He.
        The only ``background'' measurement
        at 4.95~MeV was taken with a cold, empty target.  The result here
        was consistent with zero background asymmetry.
\section{Results}
        \label{sec:results}
        Transmission asymmetries were measured for the transverse
        spin geometry at neutron energies of 1.94, 3.65, 4.95 and 7.46~MeV\@.
        The results are given in Table~\ref{tab:transasym} which
        includes the corresponding values of the beam and target 
        polarizations.  The errors associated with these polarizations are
        typically $\Delta P_n/P_n = 6$\% and
        $\Delta P_t/P_t = 2$\%.  The uncertainty in $P_n$ is dominated
        by the uncertainty in the $^{12}$C($\vec{p},p$)$^{12}$C and
        $^{12}$C($\vec{d},p_0$)$^{13}$C polarimeter analyzing powers.
        Results of the measurements conducted with
        the target cell filled with cold, liquid
        $^3$He (3\% polarization), as well as an empty sample
        container at both warm (1~K) and cold ($\sim$15~mK) temperatures
        are also included in Table~\ref{tab:transasym}.

        Transmission asymmetries for the longitudinal
        geometry were measured at neutron energies 3.65, 4.95, and
        7.46 MeV\@. A measurement at 1.94 MeV was not attempted
        because the longitudinal polarization-transfer coefficient
        for the $^3$H($\vec p$,$\vec n$)$^3$He reaction was expected
        to be too small to produce a useful asymmetry result
        \cite{Jar74}.  The transmission
        asymmetries at the three higher energies are listed in
        Table~\ref{tab:longasym} along with their corresponding
        beam and target polarizations.  Here again 
        $\Delta P_n/P_n = 6$\% and $\Delta P_t/P_t=2\%$.  
        At all three energies the
        longitudinal asymmetries were considerably smaller than the
        corresponding transverse asymmetries because the neutron
        polarizations were lower.  Not only did the polarized ion 
        source produce lower charged-particle polarizations during 
        the longitudinal measurements, but the longitudinal transfer
        coefficients $K^{z'}_z$ are typically smaller than their transverse
        counterparts $K^{y'}_y$.

        The values of $\Delta\sigma_T$ and $\Delta\sigma_L$
        extracted from the transmission asymmetries 
        are given in Table \ref{tab:dsres}.
        In all but one case, the background asymmetry was taken to be 
        the warm, liquid measurement listed in either 
        Table \ref{tab:transasym} or Table \ref{tab:longasym}.  The
        cold empty measurement at 4.95~MeV is used for the
        background correction to $\Delta\sigma_T$ at that energy.
        Both statistical and systematic uncertainties are given.
        The former reflect the counting statistics associated with 
        a measurement of the transmission asymmetry.  The systematic 
        uncertainties are based on uncertainties in beam and
        target polarizations, as well as target thickness.

\section{Comparison to Phase-Shift Predictions}
    \label{sec:comparison} 
        The $\Delta\sigma_T$ and $\Delta\sigma_L$ results are plotted in 
        Figures~\ref{fig:dstresults} and \ref{fig:dslresults}, respectively.
        The error bars shown in the figures were obtained by adding
        the systematic and statistical
        uncertainties in quadrature.
        For completeness
        we include the unpolarized neutron total cross section
        $\sigma_0$ in Fig.~\ref{fig:s0results}.  Experimental
        results are represented by the ENDF/B-VI polynomial fit
        (dash-dotted line)
        \cite{Hal91} to the data of~\cite{Bat59,Gou73,Hae83}.
        Included in Figures \ref{fig:dstresults}--\ref{fig:s0results} are
        predictions of $\Delta\sigma_T$, $\Delta\sigma_L$, and $\sigma_0$
        calculated using $n$-$^3$He phase shifts obtained from a 
        variety of sources.  Briefly, the phase shifts result
        from two published sets of partial-wave
        analyses (PWA) of $n$-$^3$He scattering
        data \cite{Jan88,Lis75}, a charge-independent
        $R$-matrix analysis of virtually all $A=4$ 
        scattering and reaction data below excitation energies of 30
        MeV \cite{Hal89}, and the preliminary results of a                    
        microscopic, variational calculation of the $^4$He continuum
        \cite{Hof93}.
        The latter is a multi-channel resonating group model (MCRGM)
        calculation that uses a gaussian-parameterized
        version of the Bonn meson-exchange potential \cite{Kel89} as its 
        input.  The $\Delta\sigma_L$ and $\Delta\sigma_T$
        calculations have been presented and
        discussed in greater detail in an earlier paper \cite{Kei94}.
        
        Three sets of phase shifts adequately reproduce $\sigma_0$, only
        the MCRGM values are clearly too low.  The MCRGM phases
        also produce values of $\Delta\sigma_L$ and $\Delta\sigma_T$ that
        are significantly lower than experiment.  All three cases
        can be attributed to
        insufficient $P$-wave amplitudes, especially the $^3\!P_2$
        partial wave.  Between neutron energies of 2 and 5 MeV,
        the $^3\!P_2$ wave is the dominant
        partial wave in all four sets of phase 
        shifts, although there is considerable discrepancy as to its
        strength.  The $^3\!P_2$ wave of both the MCRGM and the Lisowski PWA
        are nearly identical to one another, but they are considerably
        smaller than those of the $R$-matrix or Jany PWA analyses.  
        Consequently these two sets of phase shifts predict the 
        lowest values of both $\Delta\sigma_L$ and $\Delta\sigma_T$.
        To correctly reproduce $\sigma_0$, the Lisowski PWA
        compensates for its relatively small $^3\!P_2$ wave with
        unusually large $D$ waves,  particularly $^1\!D_2$.  
        Since spin-singlet
        states can only be formed when the beam and target spins are
        antiparallel to one another, Lisowski {\it et al.'s\/} 
        large $^1\!D_2$ amplitude
        further lessen their predictions of $\Delta\sigma_L$ and 
        $\Delta\sigma_T$.
       
        On the other hand, the Jany PWA possesses the largest
        $^3\!P_2$ wave, ascribing over 60\% of the total (unpolarized)
        cross section at 2~MeV to this particular wave.  Likewise
        the Jany PWA predicts $\Delta\sigma_L$ and $\Delta\sigma_T$
        values that are slightly higher than experiment.
        
        The $R$-matrix phase shifts reproduce $\Delta\sigma_T$ at all four
        energies.  The $R$-matrix prediction of $\Delta\sigma_L$
        comes closest to the measured values, although it is
        higher than experiment at 3.65~MeV\@.
        We see from Fig.~\ref{fig:s0results} that the
        $R$-matrix also overpredicts the unpolarized total cross section
        $\sigma_0$ by nearly 200~mb at this energy.  One possible explanation
        is the $^3\!P_1$ partial wave which, in the 
        $R$-matrix analysis, is
        much larger at 3.65~MeV than in the other three analyses.  While
        $\Delta\sigma_T$ is completely insensitive to this partial
        wave, $\Delta\sigma_L$ is extremely so.  If the $R$-matrix
        200~mb overprediction of $\sigma_0$ is completely attributed
        to the $^3\!P_1$ partial wave, it should likewise overpredict
        $\Delta\sigma_L$ by 400~mb (see Eq.~\ref{DsigL}).
        The experimental result of $\Delta\sigma_L$ at 3.65~MeV
        is consistent with this conclusion.

        The primary sources of the $^3\!P_1$ partial wave
        are a pair of $1^-$ resonances at 23.6~MeV ($T=1$) and
        24.2~MeV ($T=0$).  According to the $R$-matrix analysis,
        both of these excited states are predominately spin-triplet
        in the nucleon-trinucleon channels.  The $\Delta\sigma_L$
        result at 3.65~MeV, in conjunction with the unpolarized
        neutron total cross section at that energy, may indicate that
        the resonance parameters associated with one or both of the
        $1^-$ levels are in need of slight adjustment.
\section{Summary and Conclusions}
   \label{sec:conclusion}
        We have reported measurements of the polarized
        neutron---polarized $^3$He total cross section.
        A cryogenically-polarized target consisting of nearly 
        1/2 mole of solid $^3$He has been developed
        for these measurements.  It is the largest sample of 
        polarized $^3$He
        yet utilized in a nuclear physics experiment.  It is 
        particularly well suited for neutral beams such as 
        neutrons or (real) photons, where the
        sources of beam-related heating are minimal.

        Measurements of the longitudinal and
        transverse total cross-section differences
        $\Delta\sigma_L$ and $\Delta\sigma_T$ were performed 
        for incident neutron energies 2--8~MeV\@.  The results are 
        reproduced by phase shifts obtained in a recent
        $R$-matrix analysis of $A=4$ scattering and reaction data.
        As such they provide additional
        support to the $^4$He level scheme resulting from that
        analysis.  However, the measurement of $\Delta\sigma_L$ at 3.65~MeV,
        in conjunction with the unpolarized
        neutron total cross section at that energy, may indicate that
        a modification of the $R$-matrix $^3\!P_1$ partial wave
        is necessary.

        None of the other three sets of phase shifts considered here are
        able to reproduce both the present data and previous
        measurements of the unpolarized neutron total cross section.
        In all instances we are able to trace the discrepancies
        to only one or two partial waves.  In particular, we find
        clear evidence that the $^1\!D_2$ phase shift reported
        by Lisowski {\it et al.\/} is too large.

        In the future we plan
        to extend the present measurements to lower energies where
        the number of partial waves involved in the scattering and
        reaction processes is limited to two or three.
        In such circumstances it is possible
        to uniquely extract all pertinent phase-shift 
        information.
\section{Acknowledgements}
                This work was supported in part by the US Department
                of Energy,
                Office  of High Energy and Nuclear Physics, under contracts
                DE-FG05-88-ER40441 and DE-FG05-91-ER40619.

        \pagebreak
        \begin{figure}
           \caption[target]{The polarized solid $^3$He target cell.
                The $^3$He sample space is indicated by the gray shaded 
                portion.  The interior dimensions of the target are
                given on the right.}
            \label{fig:target}
        \end{figure}%
        \begin{figure}
           \caption[poltar]{Diagram of the experimental apparatus
             showing the neutron-production target, polarized target, 
             collimation, and neutron detectors.}
            \label{fig:poltar}
        \end{figure}%
        \begin{figure}
           \caption[dstresults]{Measured values of $\Delta\sigma_T$ 
             (triangles).  Error bars indicate systematic and
             statistical uncertainties added in quadrature.  Also
             shown are phase-shift predictions of $\Delta\sigma_T$:
             $R$ matrix (solid line), MCRGM (dashed line),
             Jany PWA (diamonds) and Lisowski PWA (circles).}
            \label{fig:dstresults}
        \end{figure}%
        \begin{figure}
           \caption[dslresults]{Measured values of $\Delta\sigma_L$.
             Symbols same as Fig.~\ref{fig:dstresults}.}
            \label{fig:dslresults}
        \end{figure}
       \begin{figure}
           \caption[s0results]{Phase-shift calculations of the
             unpolarized neutron total cross section
             $\sigma_0$.  Symbols same as 
             Fig.~\ref{fig:dstresults}.  The ENDF/B-VI polynomial
             fit \cite{Hal91} to the experimental data of
             \cite{Bat59,Gou73,Hae83}
             is also given (dash-dotted line).}
            \label{fig:s0results}
        \end{figure}

%
%
%       \pagebreak
%              
       \begin{table}
          \caption{Results of transverse neutron-transmission
          asymmetries.  Here $P_t$ and $P_n$ are the $^3$He and
          neutron polarizations, respectively.  The uncertainty
          quoted for each transmission asymmetry $\bar{\varepsilon}$
          reflects counting statistics, while $\sigma_{\bar{\varepsilon}}$
          is the reduced standard deviation for all the eight-step
          sequences corresponding to a measurement of
          $\bar{\varepsilon}$.  In the case of the 
          $^2$H($\vec d$,$\vec n$)$^3$He measurements, $\varepsilon_m$
          is the asymmetry observed in the neutron monitor detector.}
          \label{tab:transasym}
\begin{tabular}{clccrcr}
$E_n$ (MeV)     
       & Target Cell   
              & $P_t$    
                    & $P_n$ 
                           & $\bar{\varepsilon} $(10$^{-4}$)
                                 & $\sigma_{\bar{\varepsilon}}$
                                 (10$^{-4}$) & $\varepsilon_m$ (10$^{-4}$) \\
\hline
1.94    & cold solid    & 0.365  & 0.482 &$-42.94\pm1.72$ & 1.86 &---\\ 
        & cold liquid   & 0.029  & 0.482 &$-5.12\pm1.90$  & 2.62 &---\\ 
        & warm liquid   & 0.000  & 0.482 &$3.15\pm2.15$   & 2.34 &---\\
3.65    & cold solid    & 0.351  & 0.530 &$-45.52\pm2.16$ & 2.50 &---\\ 
        & cold liquid   & 0.029  & 0.500 &$-5.90\pm2.27$  & 2.58 &---\\ 
        & warm liquid   & 0.000  & 0.492 &$0.66\pm2.09$   & 2.40 &---\\
4.95    & cold solid    & 0.307  & 0.521 &$-30.22\pm1.05$ & 1.04 
                                                           &$68.33\pm0.54$\\ 
        & cold empty    & 0.000  & 0.521 &$-0.98\pm0.98$  & 0.97 
                                                           &$58.49\pm0.53$\\
7.46    & cold solid    & 0.345  & 0.632 &$-23.57\pm0.64$ & 0.65 
                                                           &$51.71\pm0.40$\\ 
        & cold empty    & 0.000  & 0.639 &$-2.51\pm0.57$  & 0.57 
                                                           &$54.66\pm0.35$\\ 
        & warm empty    & 0.000  & 0.634 &$-3.76\pm0.64$  & 0.64 
                                                           &$66.81\pm0.39$\\ 
        & warm liquid   & 0.000  & 0.622 &$-3.17\pm0.59$  & 0.59 
                                                           &$54.73\pm0.33$\\
      \end{tabular}
\end{table}
%       \pagebreak
%              
       \begin{table}
          \caption{Results of longitudinal neutron-transmission
          asymmetries.}
          \label{tab:longasym}
\begin{tabular}{clccrcr}
$E_n$ (MeV)     
       & Target Cell   
              & $P_t$    
                    & $P_n$ 
                           & $\bar{\varepsilon} $(10$^{-4}$)
                                 & $\sigma_{\bar{\varepsilon}}$
                                 (10$^{-4}$) & $\varepsilon_m$ (10$^{-4}$)\\
\hline
3.65    & cold solid    & 0.351  & 0.263 &$-8.65\pm1.89$ & 1.91 &---\\ 
        & warm liquid   & 0.000  & 0.265 &$1.81\pm1.92$   & 1.95 &---\\
4.95    & cold solid    & 0.352  & 0.334 &$-18.81\pm1.17$ & 1.19 
                                                       &$263.4\pm0.87$\\ 
        & warm liquid    & 0.000  & 0.372 &$2.16\pm1.19$  & 1.22 
                                                       &$250.2\pm0.75$\\
7.46    & cold solid    & 0.343  & 0.404 &$-6.34\pm0.60$ & 0.60 
                                                       &$127.0\pm0.48$\\ 
        & warm liquid   & 0.000  & 0.210 &$4.37\pm0.62$  & 0.63 
                                                       &$134.7\pm0.47$\\
      \end{tabular}
\end{table}
\begin{table}
  \caption{Results of the $\Delta\sigma_L$ and $\Delta\sigma_T$ 
    measurements.  The first
    uncertainty is systematic, the second statistical.}
  \label{tab:dsres}

                \begin{tabular}{ccc}
                        $E_{n}$ (MeV)
                                & $\Delta\sigma_{L}$ (b)
                                & $\Delta\sigma_{T}$ (b) \\
                        \hline
                        \hspace{0.5em}1.94
                                & ---
                                & $\hspace{1.25em}1.207\pm0.092
                                        \pm0.078$ \\
                        \hspace{0.5em}3.65
                                & $\hspace{1.25em}0.432\pm0.044
                                        \pm0.095$ 
                                & $\hspace{1.25em}1.145\pm0.089
                                        \pm0.089$ \\
                        \hspace{0.5em}4.95
                                & $\hspace{1.25em}0.806\pm0.068
                                        \pm0.067$
                                & $\hspace{1.25em}0.838\pm0.073
                                        \pm0.044$ \\
                        \hspace{0.5em}7.46
                                & $\hspace{1.25em}0.348\pm0.041
                                        \pm0.028$
                                & $\hspace{1.25em}0.431\pm0.035
                                        \pm0.018$
                \end{tabular}
\end{table}

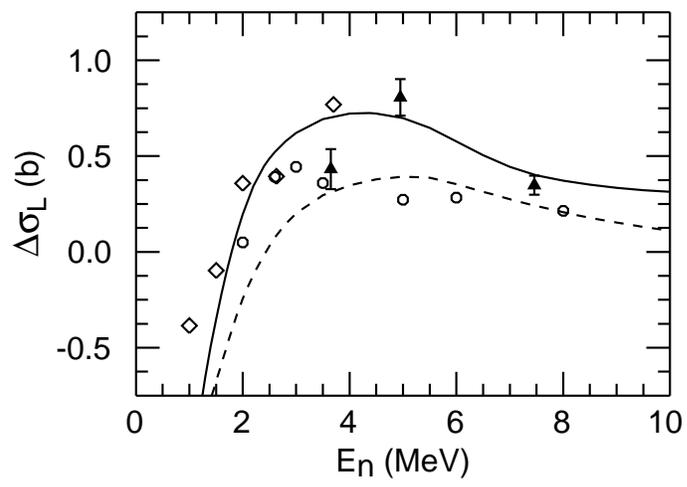
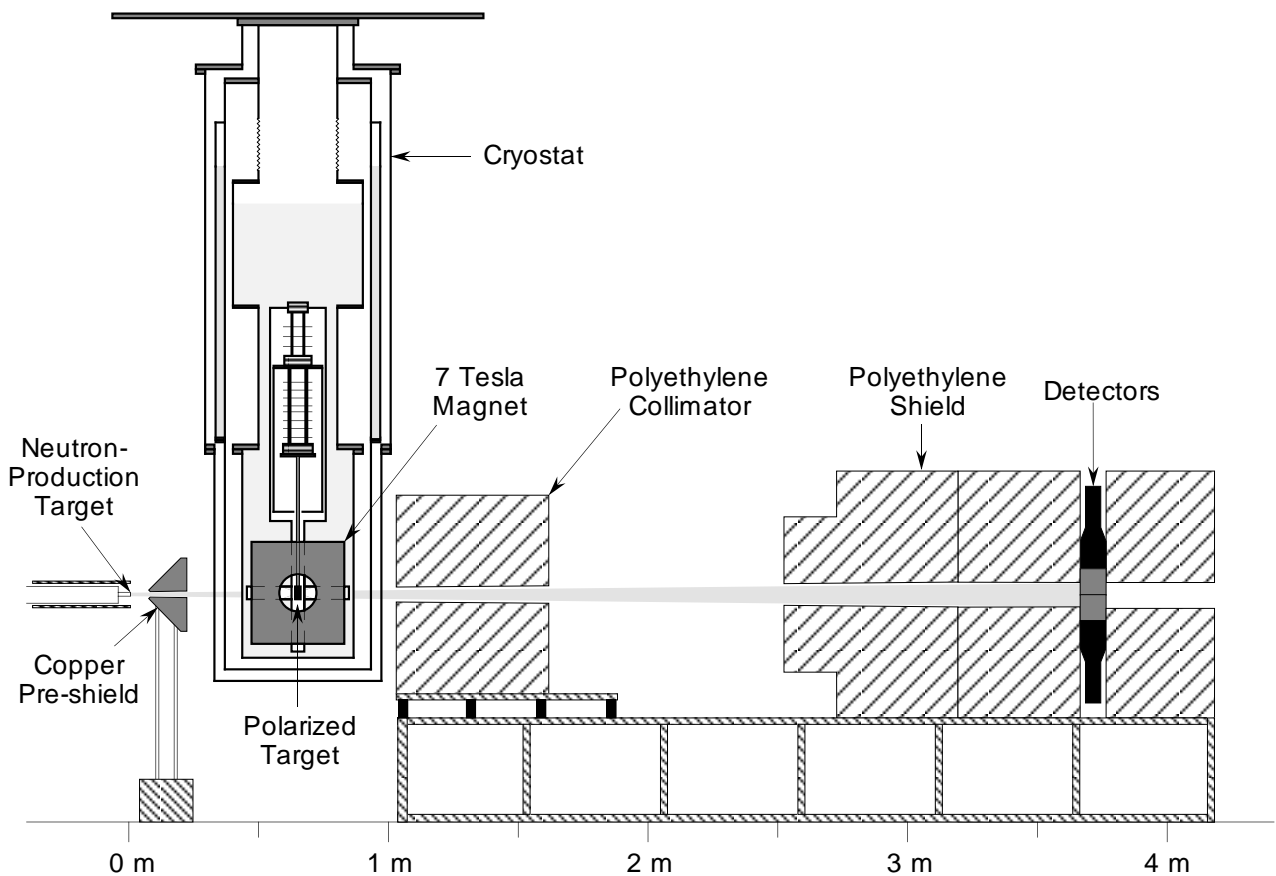
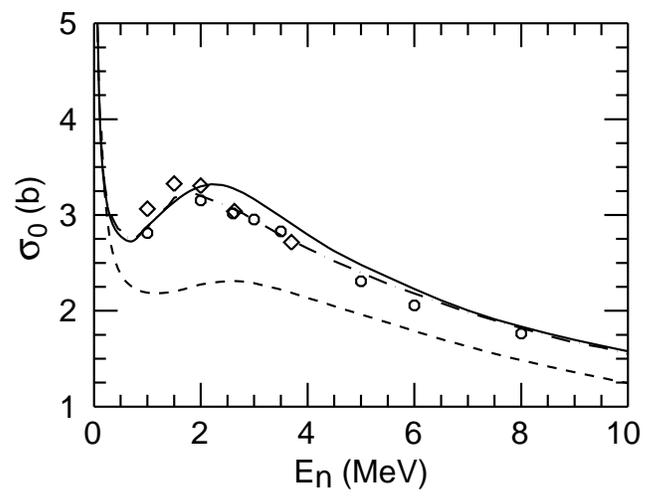
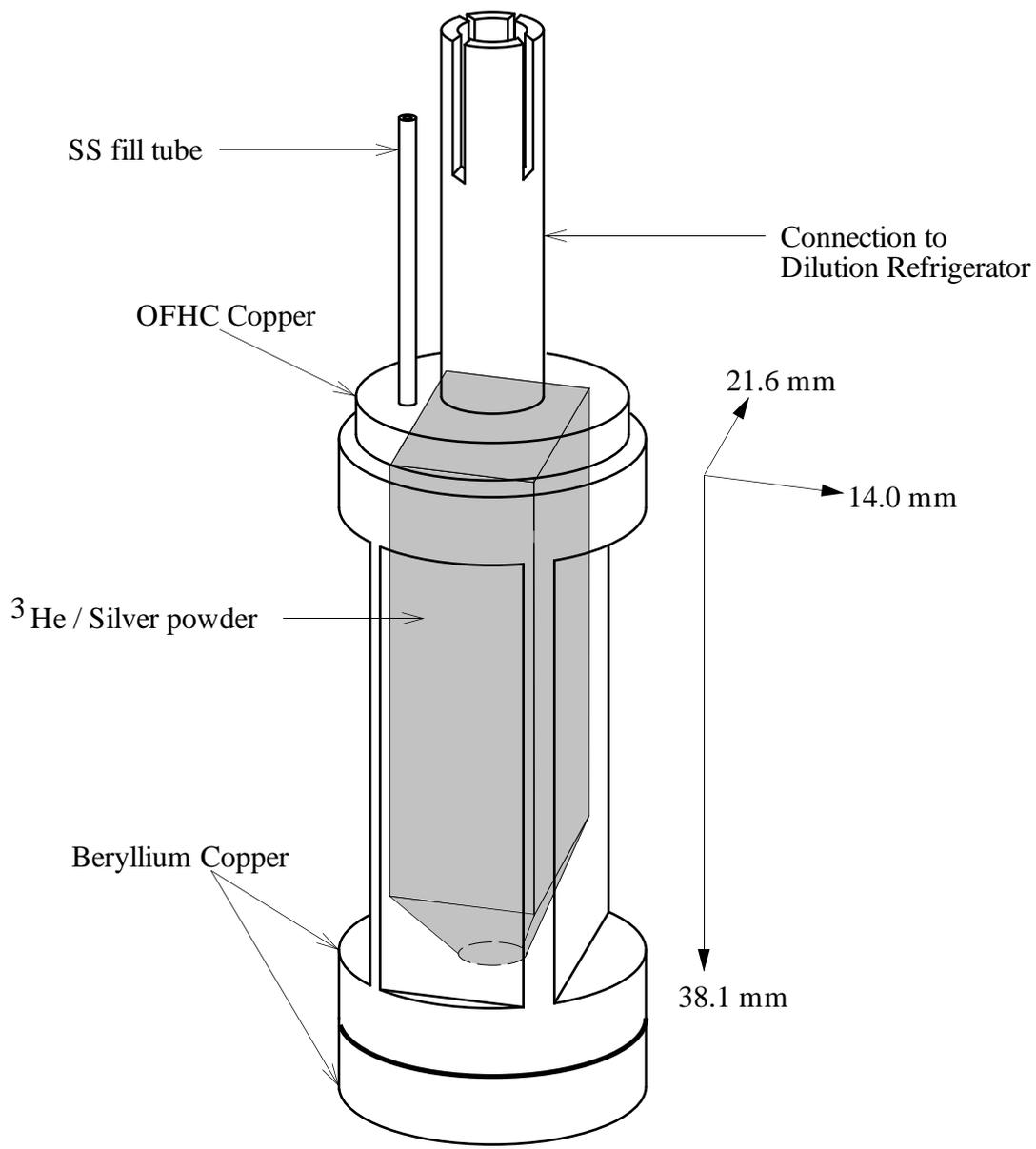
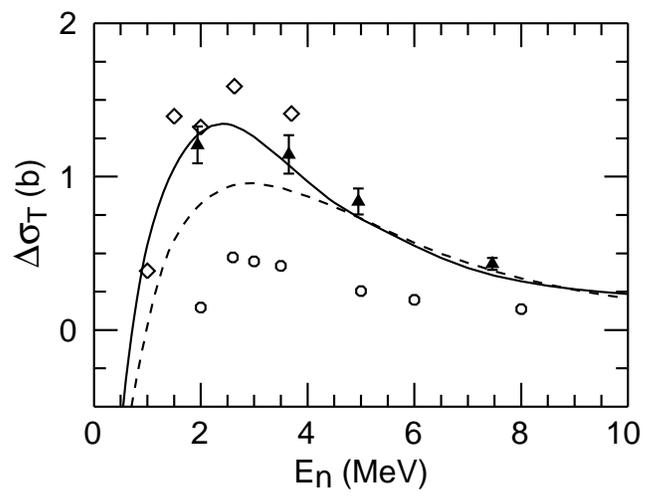
\end{document}